\documentclass{article}
\usepackage{epsfig}
\usepackage[latin2]{inputenc}
\begin{document}
\title{Quantitive and sociological analysis of blog networks
}
\author{Wiktor Bachnik, Stanislaw Szymczyk, Piotr Leszczynski, \\
Rafal Podsiadlo, Ewa Rymszewicz, Lukasz Kurylo \\
(students of Computer Science, Gda\'nsk University), \\
Danuta Makowiec\\(Institute of Theoretical Physics and Astrophysics, Gda\'nsk University),\\
Beata Bykowska\\(Institute of Philosophy and Sociology, Gda\'nsk University)}
\maketitle
\begin{abstract}
This paper examines the emerging phenomenon of blogging, using three 
different Polish blogging services as the base of the research. 
Authors show that blog networks are sharing their characteristics
with complex networks ($\gamma$ coefficients, small worlds, cliques, 
etc.). Elements of sociometric analysis were used to 
prove existence of some social structures in the blog networks.
\end{abstract}
  
\section{Introduction}

\subsection{Blog -- what is it?}

\emph{Blog}\footnote{http://www.matisse.net/files/glossary.html\#Blog}
\footnote{http://www.blogger.com/tour\_start.g} is a diary published on 
the author's website. Because the Internet is used as a medium, authors 
feel free to express their opinions and views on different subjects, 
without fear of censorship. 

\subsection{How blog networks are created?}

As \emph{blogging} becomes very popular, many internet portals offer (mostly free 
of charge) blogging facilities to their customers. That causes aggregation of blogs 
in one ``place'', and encourages building communities. Bloggers (as we call people 
who run their blogs) very often place hypertext links to their friends and colleagues 
sharing  similar views or describing similar subjects. Such connections create
what we call \emph{blog networks} which are subject of our research.

\subsection{Examined blogging services}

We examined three different Polish blogging services:

\begin{enumerate}
\item \textbf{blog.onet.pl} -- one of the most popular services, about $150,000$ registered blogs
\item \textbf{blog.gery.pl} -- moderately known service, about $15,000$ blogs
\item \textbf{jogger.pl} -- niche service, gathering mostly tech-savvy people, only around $1,500$ blogs
\end{enumerate}

It should noted that many of blogs may be abandoned by their authors and no longer updated.
They are still available however, and were taken into the account.

\section{Collecting the data}

We used standard GNU/Linux tools to automate process of collecting the data:
\begin{itemize}
\item text-mode {\it lynx} browser for downloading the content of WWW pages
\item {\it grep} for filtering out unnecessary information
\item {\it sort} for sorting the blog list
\item {\it uniq} for removing the duplicate blog list entries
\item {\it bash} shell which provided a scripting framework
\end{itemize}
Usually blogging services provide users with possibility of listing all
existing blogs. We used this feature to create a list of all bloggers for each
service. For example jogger.pl blog list has the following URL:
\begin{center}{\tt http://jogger.pl/users.php?sort=1\&start=offset}\end{center}
\noindent where {\tt offset is} the CGI parameter for specifying position in the list.
It has 100 blog links presented on each page, so it was possible to gather all
the blog links by starting from {\tt offset}=0 and increasing it by 100 in a loop
until no more blogs were presented. In each loop iteration content of the
list page was downloaded by using {\it lynx} browser in HTML source dump mode.
Then {\it grep} was used to filter out all data apart from blog URL addresses.
We found it convenient to sort the resulting list and remove duplicate
entries.

When the list was ready, content of every listed blog page was downloaded and
links to other blogs in the same service were filtered out in similar manner.
In the result, list of all outgoing connections for each blogger in the
service was created. This process was repeated for each examined blogging service.

\section{Quantitive analysis}

This section presents results of quantitive analysis performed on data collected 
from the services we examined. 

\subsection{Vertices}

The terminology we used comes from the graph theory. Each blog is represented by a 
vertex in the connection graph.

Average vertex degrees for each service:

\begin{enumerate}
\item \textbf{blog.onet.pl}: $0.8105$
\item \textbf{blog.gery.pl}: $0.5243$
\item \textbf{jogger.pl}: $0.4392$
\end{enumerate}

It's clearly seen that these graphs are very sparse. We'll try to show that the 
function of degree distribution is of power--law type: $Count(k) \propto k^{-\gamma}$, 
where $k$ represents vertex degree.

\begin{figure}[htbp]
\centering
\epsfig{figure=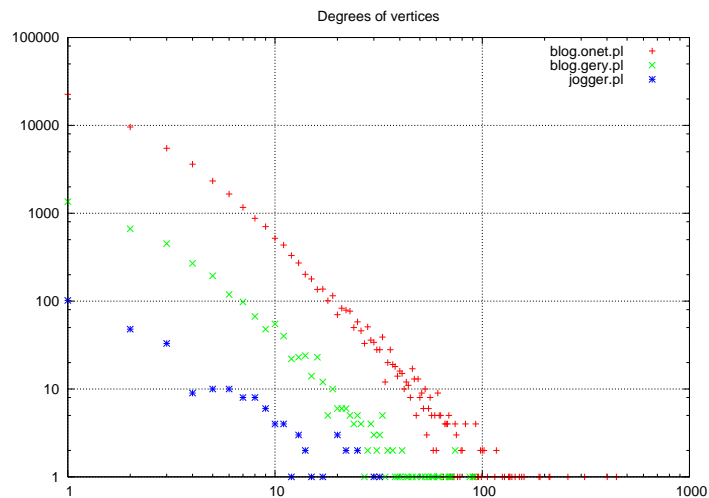,width=0.8\textwidth}
\caption{Histogram of vertex degrees: incoming and outgoing edges combined, log--log plots}
\label{degrees1}
\end{figure}
\begin{figure}[htbp]
\centering
\epsfig{figure=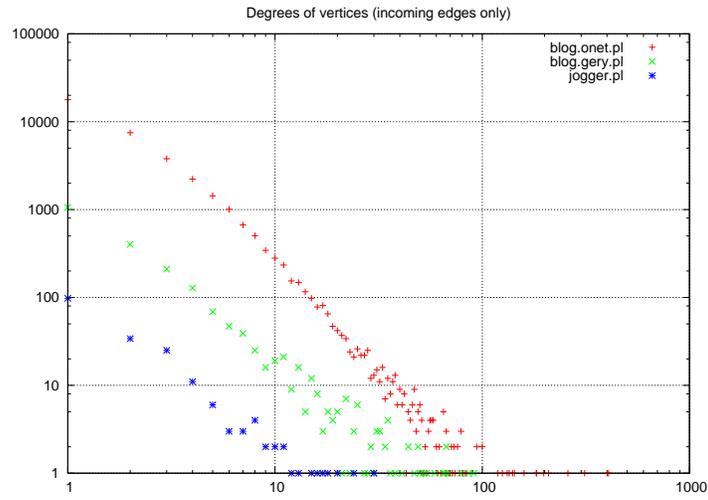,width=0.8\textwidth}
\caption{Histogram of vertex degrees: incoming edges, log--log plots}
\label{degrees2}
\end{figure}
\begin{figure}[htbp]
\centering
\epsfig{figure=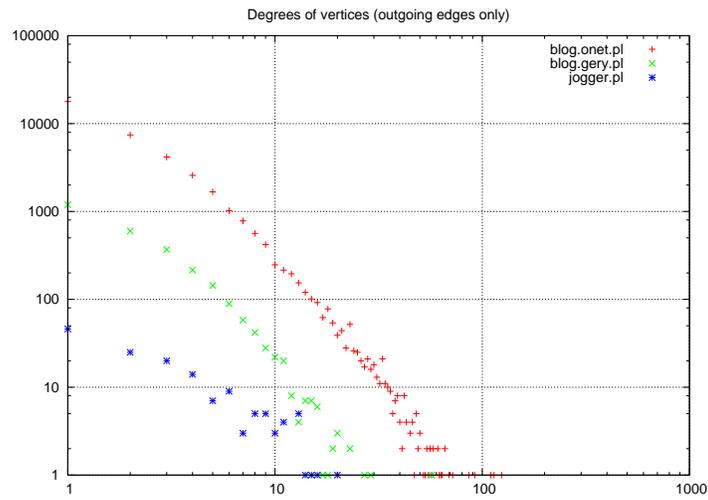,width=0.8\textwidth}
\caption{Histogram of vertex degrees: outgoing edges, log--log plots}
\label{degrees3}
\end{figure}


Histograms presented in Figures~\ref{degrees1},~\ref{degrees2} 
and~\ref{degrees3} are very similar, even though 
number of blogs in each service is different by an order of magnitude. 
That shows us that scaling is also very similar in these networks.
 
$\gamma$ coefficients of the vertices degree functions are presented 
in Table~\ref{tabgamma} below. $R^2$ represents the square of the correlation 
coefficient.

\begin{table}[!htbp]
\begin{center}
\caption{$\gamma$ coefficients of the vertices degree functions}
\label{tabgamma}
\begin{tabular}{|p{3.5cm}|c|c|c|c|c|c|}
\hline
Service & \multicolumn{2}{c}{blog.onet.pl} & \multicolumn{2}{c}{blog.gery.pl} & \multicolumn{2}{c}{jogger.pl}\\
\hline
 & $\gamma$ & $R^2$ & $\gamma$ & $R^2$ & $\gamma$ & $R^2$ \\
\hline
Outgoing edges & $2.96$ & $0.97$ & $3.00$ & $0.96$ & $2.14$ & $0.91$ \\
\hline
Incoming edges & $2.68$ & $0.97$ & $2.25$ & $0.93$  & $2.24$ & $0.95$ \\
\hline
Incoming and outgoing edges combined & $2.70$ & $0.97$ & $2.38$ & $0.96$ & $2.05$ & $0.92$ \\
\hline
\end{tabular}
\end{center}
\end{table}
						   
Vertices with maximal degrees are listed in Table~\ref{tabmaxvertices}.

\begin{table}[!htbp]
\begin{center}
\caption{Vertices with maximal degrees}
\label{tabmaxvertices}
\begin{tabular}{|p{2.7cm}|c|c|c|c|c|c|}
\hline
Service & \multicolumn{2}{c}{blog.onet.pl} & \multicolumn{2}{c}{blog.gery.pl} & \multicolumn{2}{c}{jogger.pl}\\
\hline
 & Name & Deg. & Name & Deg. & Name & Deg. \\
\hline
Outgoing edges & zycielily  & 407  & martus  & 91 & jpc & 30  \\
\hline
Incoming edges & blizniaczki777 & 124  & www\footnotemark & 57 & siwa & 20\\
\hline
Incoming and outgoing edges combined & zycielily  & 444 & martus & 91  & marcoos & 32 \\
\hline
\end{tabular}
\end{center}
\end{table}
\footnotetext{home page of the service}

\subsection{Average path length}

Average path lengths for each service are presented in Table~\ref{tabpaths}.
Standard deviation is represented by the $\sigma$ symbol.

\begin{table}[!htbp]
\begin{center}
\caption{Average path lengths}
\label{tabpaths}
\begin{tabular}{|p{3.5cm}|c|c|c|}
\hline
Service & blog.onet.pl & blog.gery.pl & jogger.pl\\
\hline
Average path length & 7.60 & 6.76  & 3.78 \\
\hline
$\sigma$ & 3.46 & 3.74& 2.64 \\
\hline
\end{tabular}
\end{center}
\end{table}

\begin{figure}[htbp]
\centering
\epsfig{figure=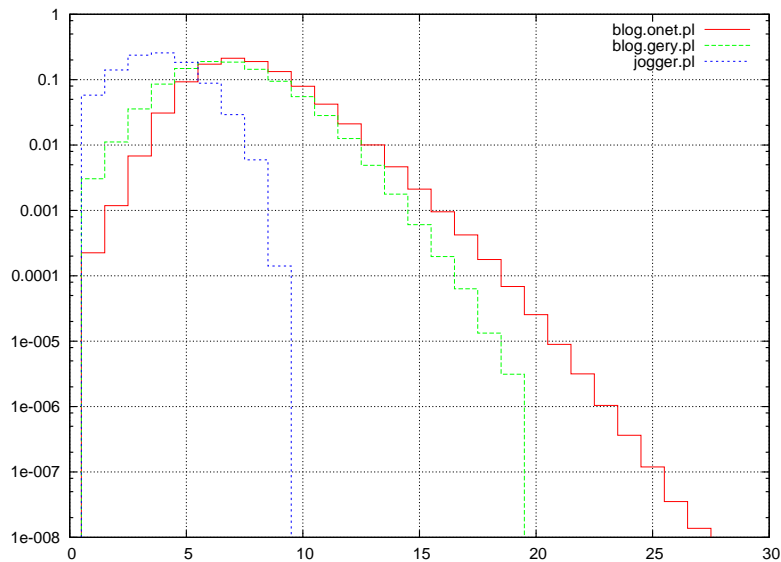,width=0.9\textwidth}
\caption{Histogram of path lengths in each service, log plot}
\label{pathshistogram}
\end{figure}

\subsection{Cliques}

Two different kinds of connections between the vertices are distinguished -- 
\textit{weak} (idols and fans) and \textit{strong} (friends). We call a connection
between vertices $A, B$ \textit{weak} when there's only one edge, going either 
from $A$ to $B$ or $B$ to $A$. That means that only one blog links to the other, 
which resembles relationship between fan and his idol. On the other hand, connection 
is called \textit{strong} when two edges between $A$ and $B$ can be found. First goes 
from $A$ to $B$ and the other from $B$ to $A$. If we assume that linking to somebody's 
blog means liking that person, then such relation means that $A$ and $B$ are friends 
as they like each other.

We also measured average cliquity for each service. Cliquity $c_i$ represents 
``completeness'' of the neighbourhood of vertex $i$ \cite{barabasi, dorogovtsev}, i.e.  
$c_i$ is $1$ in case of a complete subgraph, $0$ when a vertex is isolated. 

\begin{figure}[htbp]
\centering
\epsfig{figure=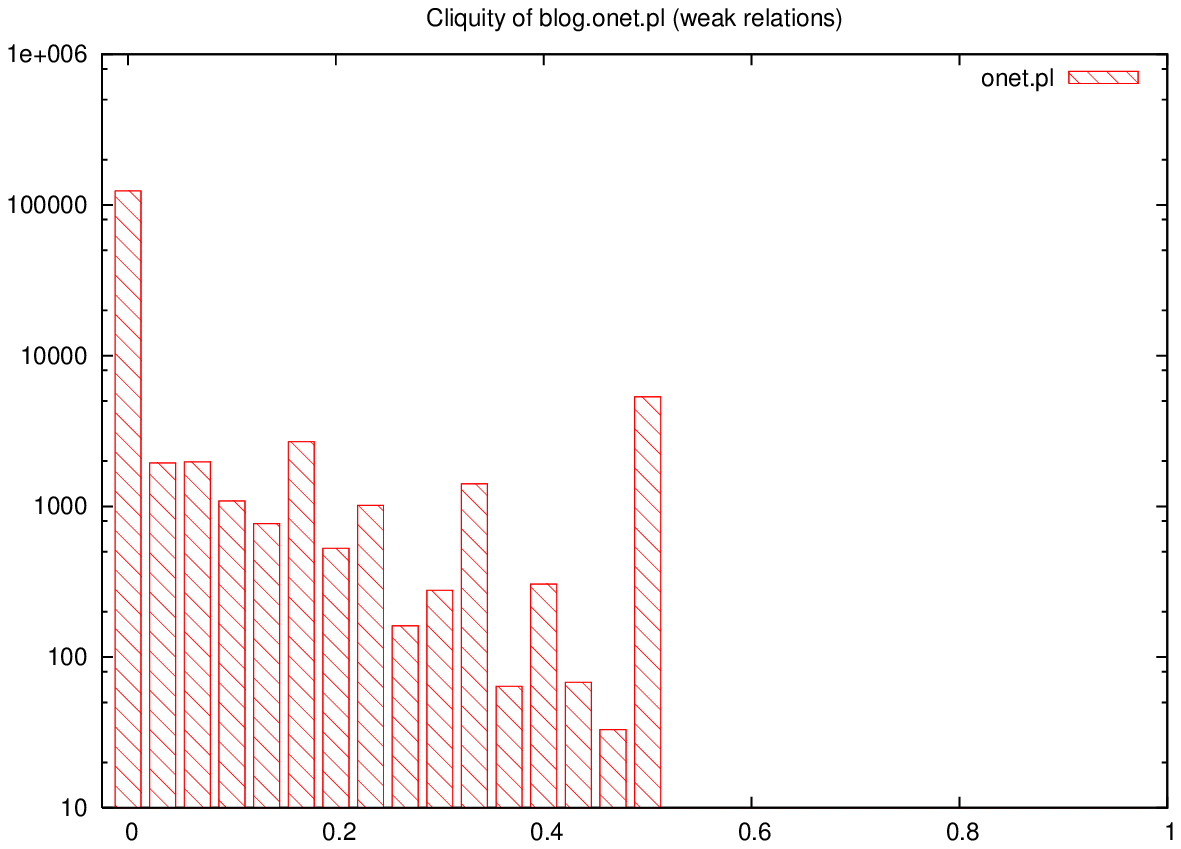,width=0.8\textwidth}
\caption{Histogram of cliquity for {\it blog.onet.pl}, weak relations, log plot}
\label{cliquesweak1}
\end{figure}

\begin{figure}[htbp]
\centering
\epsfig{figure=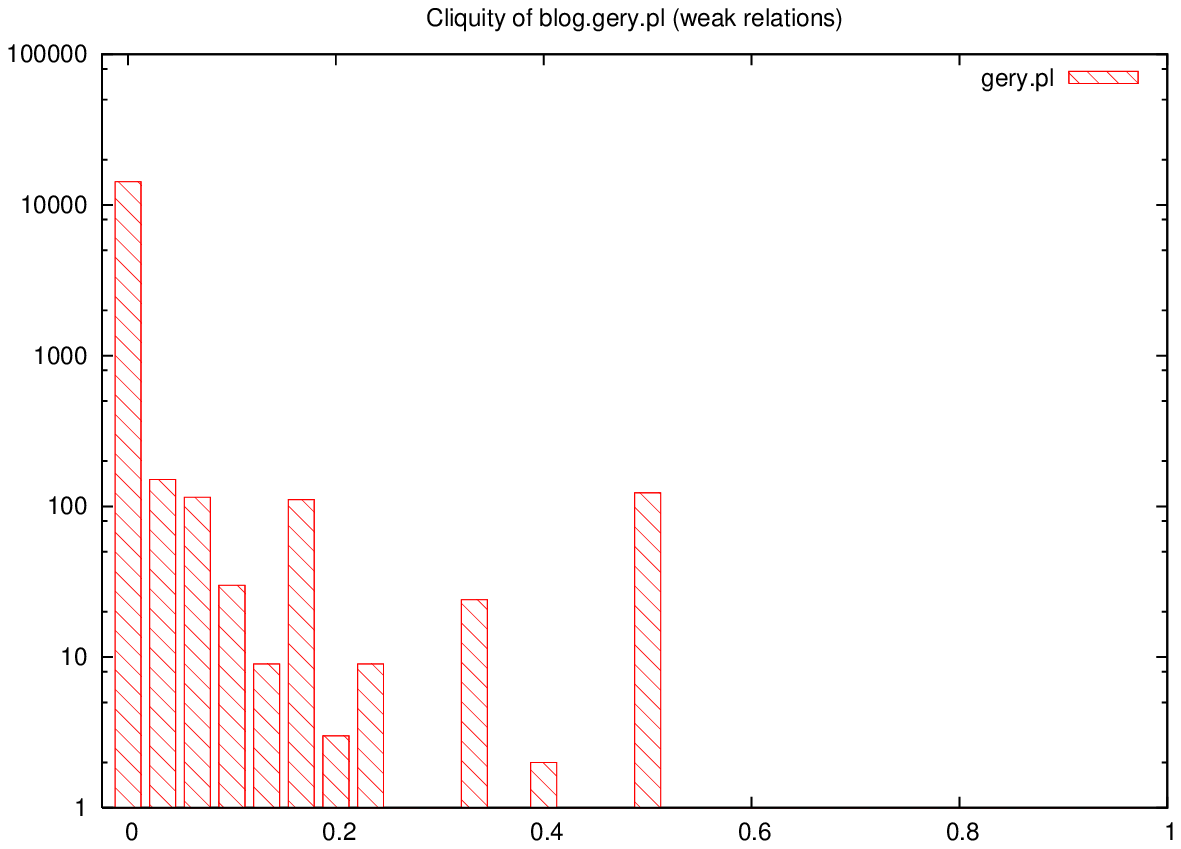,width=0.8\textwidth}
\caption{Histogram of cliquity for {\it blog.gery.pl}, weak relations, log plot}
\label{cliquesweak2}
\end{figure}

\begin{figure}[htbp]
\centering
\epsfig{figure=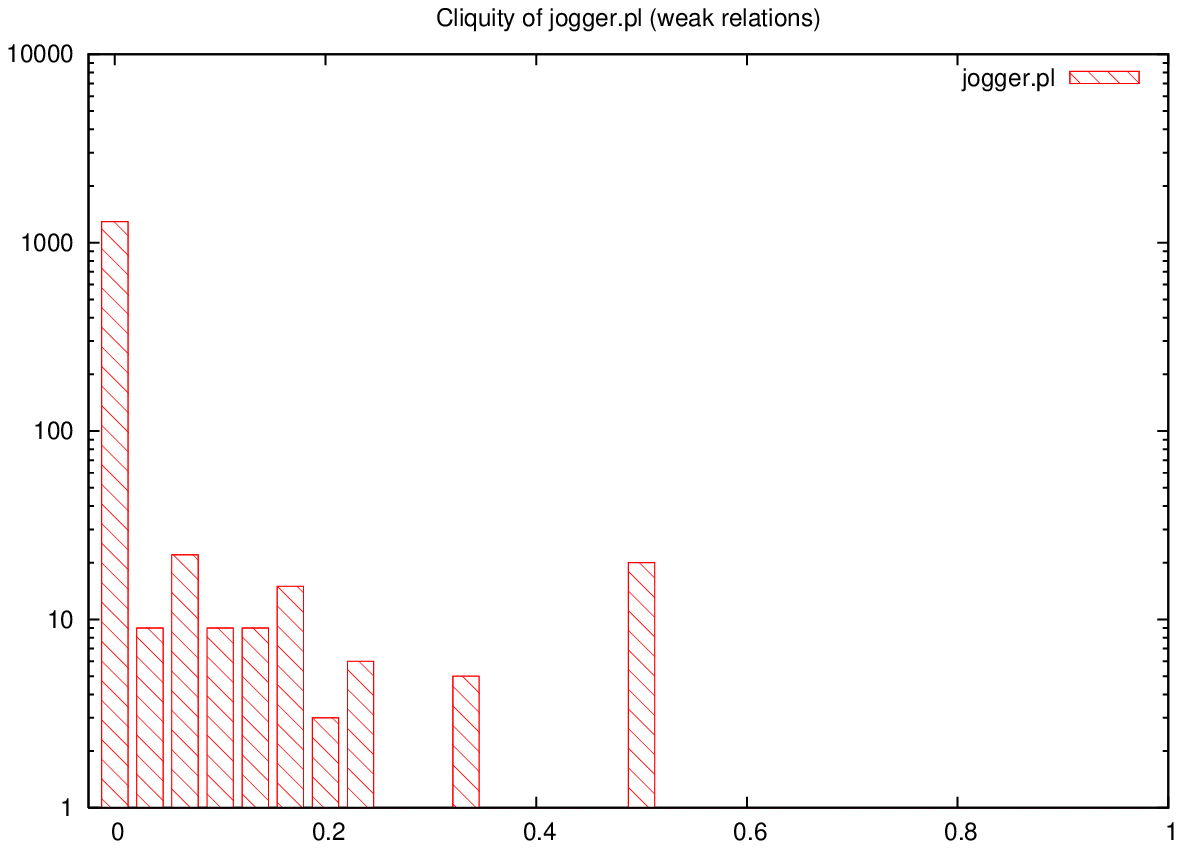,width=0.8\textwidth}
\caption{Histogram of cliquity for {\it jogger.pl}, weak relations, log plot}
\label{cliquesweak3}
\end{figure}

\begin{figure}[htbp]
\centering
\epsfig{figure=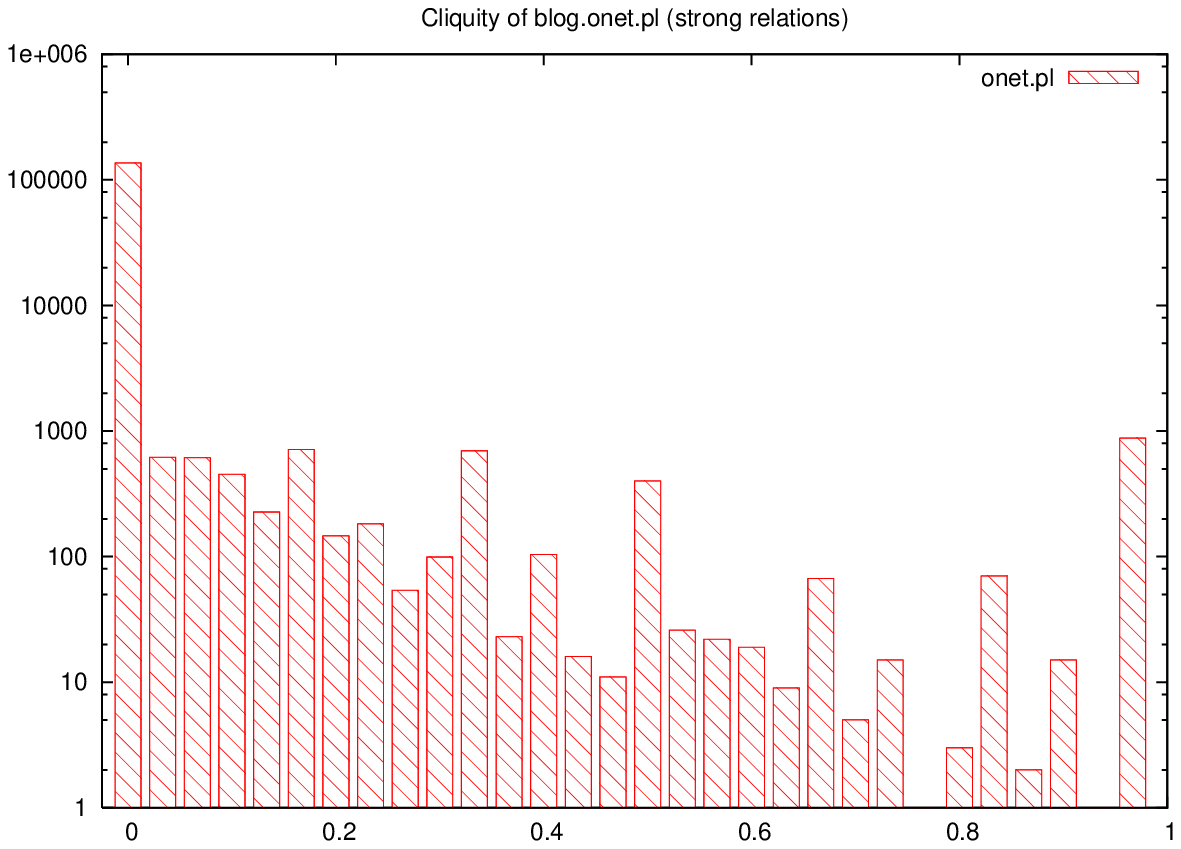,width=0.8\textwidth}
\caption{Histogram of cliquity for {\it blog.onet.pl}, strong relations, log plot}
\label{cliquesstrong1}
\end{figure}

\begin{figure}[htbp]
\centering
\epsfig{figure=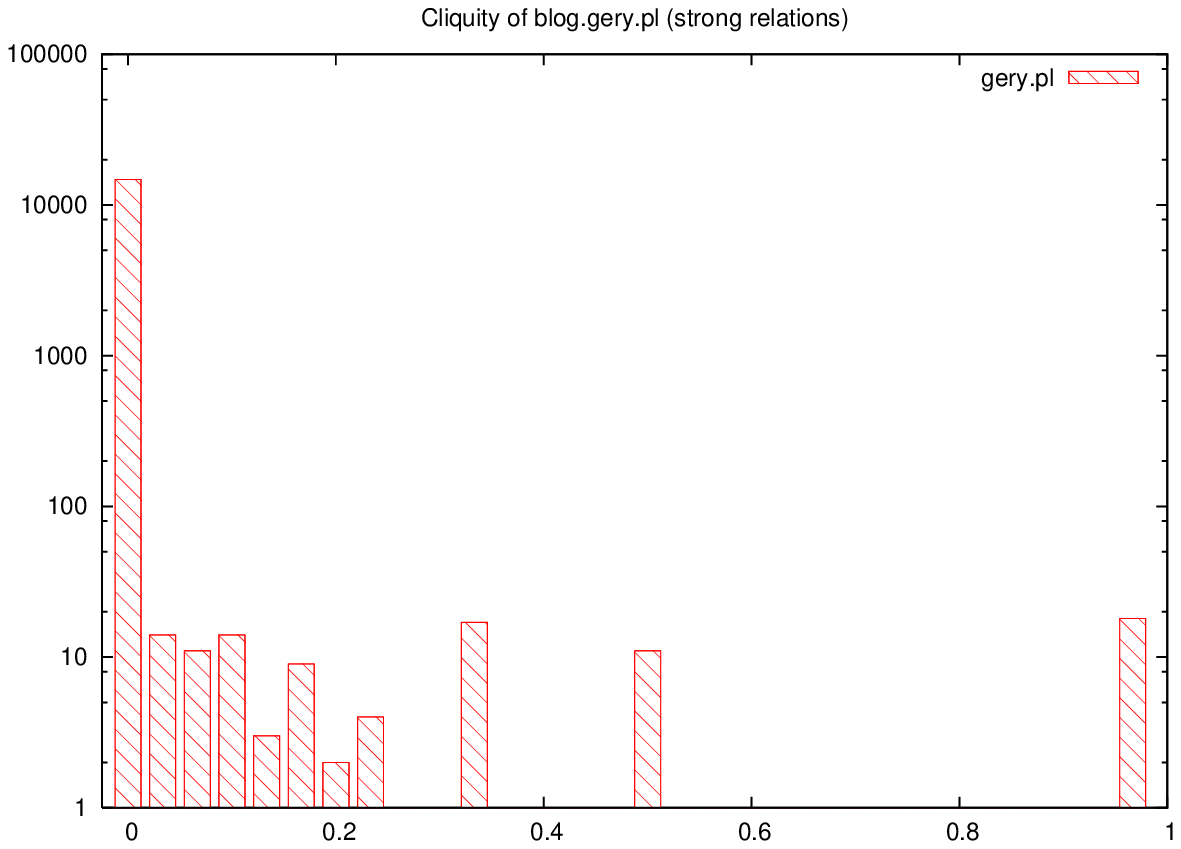,width=0.8\textwidth}
\caption{Histogram of cliquity for {\it blog.gery.pl}, strong relations, log plot}
\label{cliquesstrong2}
\end{figure}

\begin{figure}[htbp]
\centering
\epsfig{figure=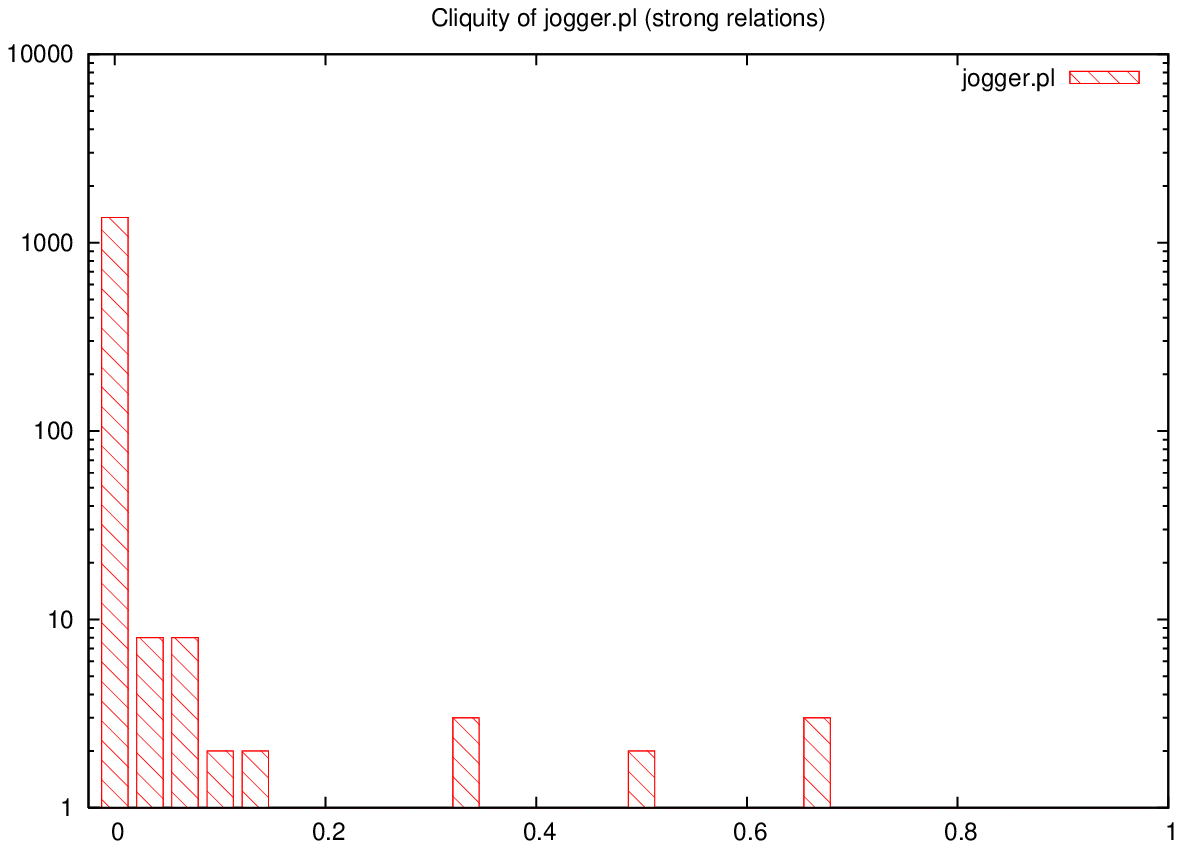,width=0.8\textwidth}
\caption{Histogram of cliquity for {\it jogger.pl}, strong relations, log plot}
\label{cliquesstrong3}
\end{figure}

Average cliquities for each service are presented in Table~\ref{tabcliques}. 
Figures~\ref{cliquesweak1}--\ref{cliquesweak3} and \ref{cliquesstrong1}--\ref{cliquesstrong3} 
show histograms of vertex cliquities for each examined service, \textit{weak} 
and \textit{strong} connections respectively. Overdominance of isolated vertices is evident. 
When \textit{strong} connections are considered, full subgraphs can be observed in larger
services.

\begin{table}[!htbp]
\begin{center}
\caption{Average cliquities for each service}
\label{tabcliques}
\begin{tabular}{|p{3.5cm}|c|c|c|c|c|c|}
\hline
Service & \multicolumn{2}{c}{blog.onet.pl} & \multicolumn{2}{c}{blog.gery.pl} & \multicolumn{2}{c}{jogger.pl}\\
\hline
 & $c$ & $\sigma$ & $c$ & $\sigma$ & $c$ & $\sigma$ \\
\hline
Weak relations & $0.067$ & $0.107$ & $0.015$ & $0.050$ & $0.030$ & $0.068$ \\
\hline
Strong relations & $0.013$ & $0.091$ & $0.002$ & $0.039$  & $0.004$ & $0.046$ \\
\hline
\end{tabular}
\end{center}
\end{table}



\section{Sociometric analysis}
Connected graph is a graph in which every two vertices are connected with a path. 
Two subgraph groups have been generated: strong relationship graphs --- when one 
blog is referring to another, the other mutually referring to the 
first one (``friends'') and weak relationship graphs --- where references are 
not mutual.      

\begin{figure}
\includegraphics[width=1\textwidth]{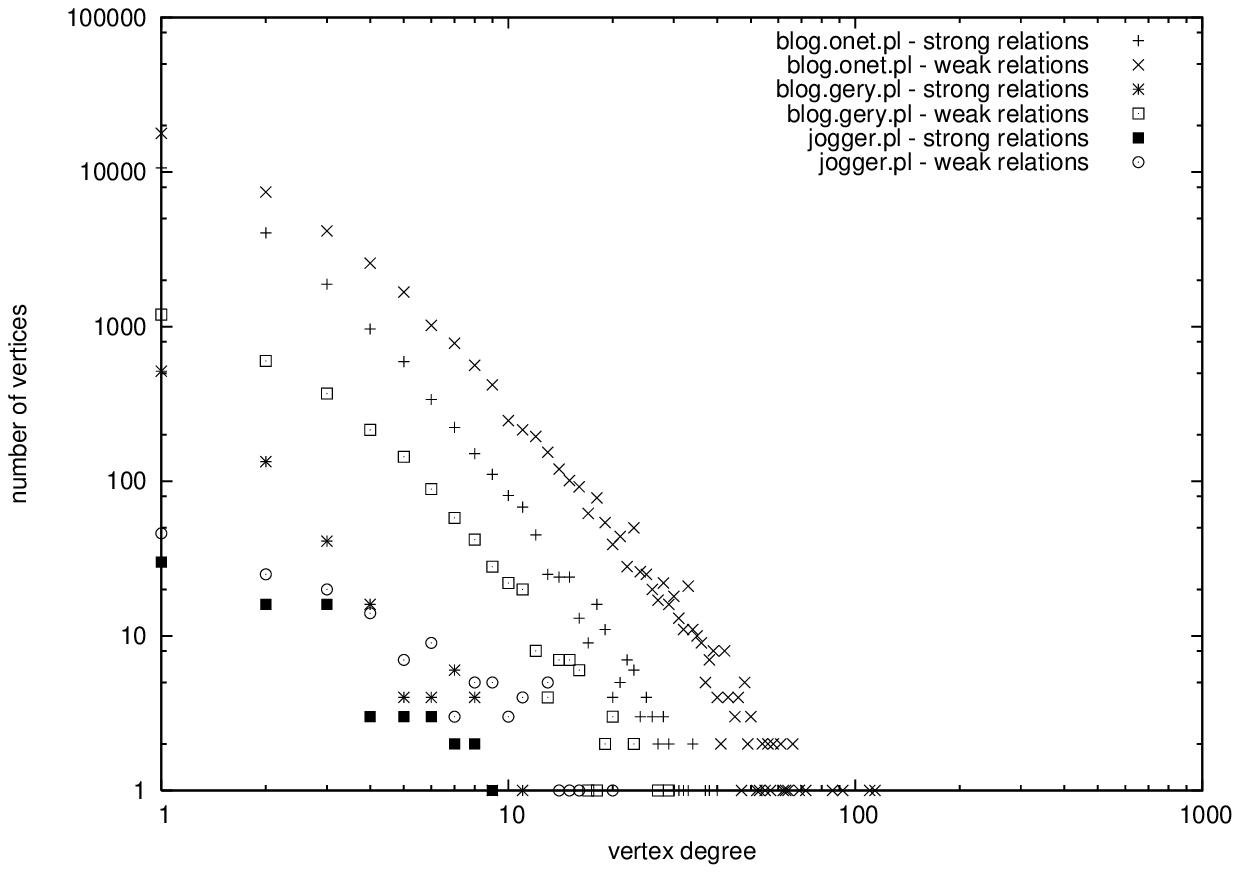}
\caption{Histogram of vertex degrees.}
\label{rys1}
\end{figure}

Frequencies of vertex degrees depending on the type 
of relationship are shown in Fig.~\ref{rys1}.
The number of isolated persons was established (no references to 
other blogs on their pages). 
The result is given in Table~\ref{tab1}. 
\begin{table}
\caption{Number of isolated users and blogs in surveyed services}
\label{tab1}
\begin{tabular}{|c|c|c|c|}
\hline
Portal & jogger.pl & blog.gery.pl & blog.onet.pl \\
\hline
Number of users & 1391 & 14861 & 141755 \\
\hline
Number of isolated blogs & 1315 & 14135 & 122412 \\
\hline
Percent of isolated blogs & 94.5\% & 95.1\% & 86.3\% \\
\hline
Percent of not isolated blogs & 5.5\% & 4.9\% & 13.7\% \\
\hline
Average number of users & 9.5 & 3.24 & 5 \\
\hline
Number of strong subgraphs & 8 & 224 & 3797 \\
\hline
\end{tabular}
\end{table}
Having given number of isolated persons from particular blog service, it is 
possible to establish group integration index. 
The integration index is calculated with the following method \cite{a3}:

$$IG = \frac{1}{Number\:of\:isolated\:persons}$$
        
These are respectively:
$IGgery = 7.8715*10^{-5}$, $IGonet = 9.5524*10^{-6}$.

As a result of computer - aided calculations we have been able to determine 
the number of blog pairs for 
blog.onet.pl and blog.gery.pl services where authors chose each other mutually 
(placed links in their weblogs). 
For blog.gery.pl this was 554 of total 14861; in case of blog.onet.pl this value 
reached 21160 of total number of 
141755 weblogs. 
Connection index is given by formula \cite{a3}:
$$SG = \frac{Number\:of\:pairs\:with\:mutual\:choices}{C^N_2}$$
Consequently, connection indices for these blogs are respectively:
$SGgery = 5.0173*10^{-6}$, $SGonet = 2.106*10^{-6}$.
Notice that despite a tenfold population difference between the two 
services, connection indices differ only about 2 times.

\begin{figure}
\centering
\includegraphics[angle=90, width=0.25\textwidth]{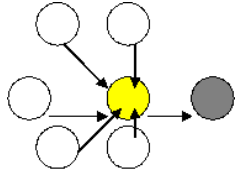}
\caption{{\it Idol} and {\it eminence grise}}
\label{r4}
\end{figure}

\textit{Idol} is a sociometric structure which describes person who got the large 
number 
of positive choices, though making small number of choices by itself 
(that means that it has small positive expansion)\cite{sztompa, a4, a7} . 
With \textit{idol} is connected 
the person of {\it eminence grise} --- who is the person chosen by {\it idol} 
(illustrated in Fig.~\ref{r4}).

\begin{figure}
\includegraphics[angle=270, width=1\textwidth]{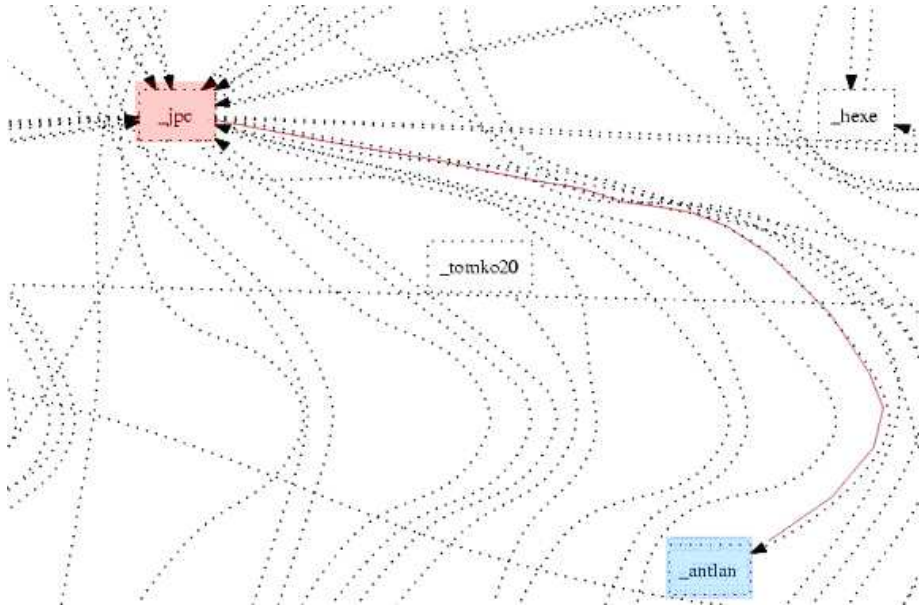}
\caption{{\it Idol} and {\it eminence grise} structure found in jogger.pl 
service network.}
\label{r5}
\end{figure}

Blog jpc (shown in Fig.~\ref{r5}) is an {\it idol} with relatively large positive 
expansion (21 choices). {\it Eminence grise} is clearly visible (blog antlan), 
and is chosen by jpc without mutuality. Text analysis suggests that 
authors of both blogs are friends from University, from the ``real'' life. The more 
experienced user (jpc) 
promotes his friend's weblog in bloggers' community. This however does not work 
very well --- although 
blog jpc was established in November 2003 and is regularly updated, 
blog antlan is an ephemeron. For the 5 months of its existence it was 
updated only once. 

\begin{figure}
\centering
\includegraphics[width=0.15\textwidth]{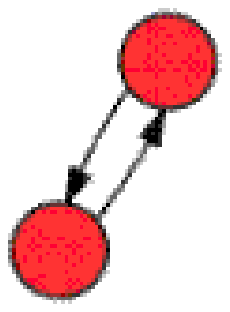}
\includegraphics[width=0.15\textwidth]{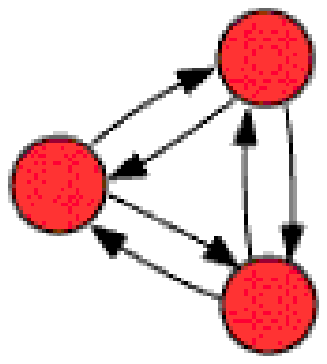}
\caption{The most popular sociometrical systems by J. 
Moreno \cite{sztompa, a2, a7} - {\it diad} and {\it triad}}
\label{diada_triada}
\end{figure}

In the Fig.~\ref{diada_triada} the basic sociometric structures are presented --- 
{\it diad} which is mutual positive choice 
between 2 persons and {\it triad} which is mutual positive choice among 3 
persons \cite{sztompa, a5, a6}.

\begin{figure}
\includegraphics[width=0.3\textwidth]{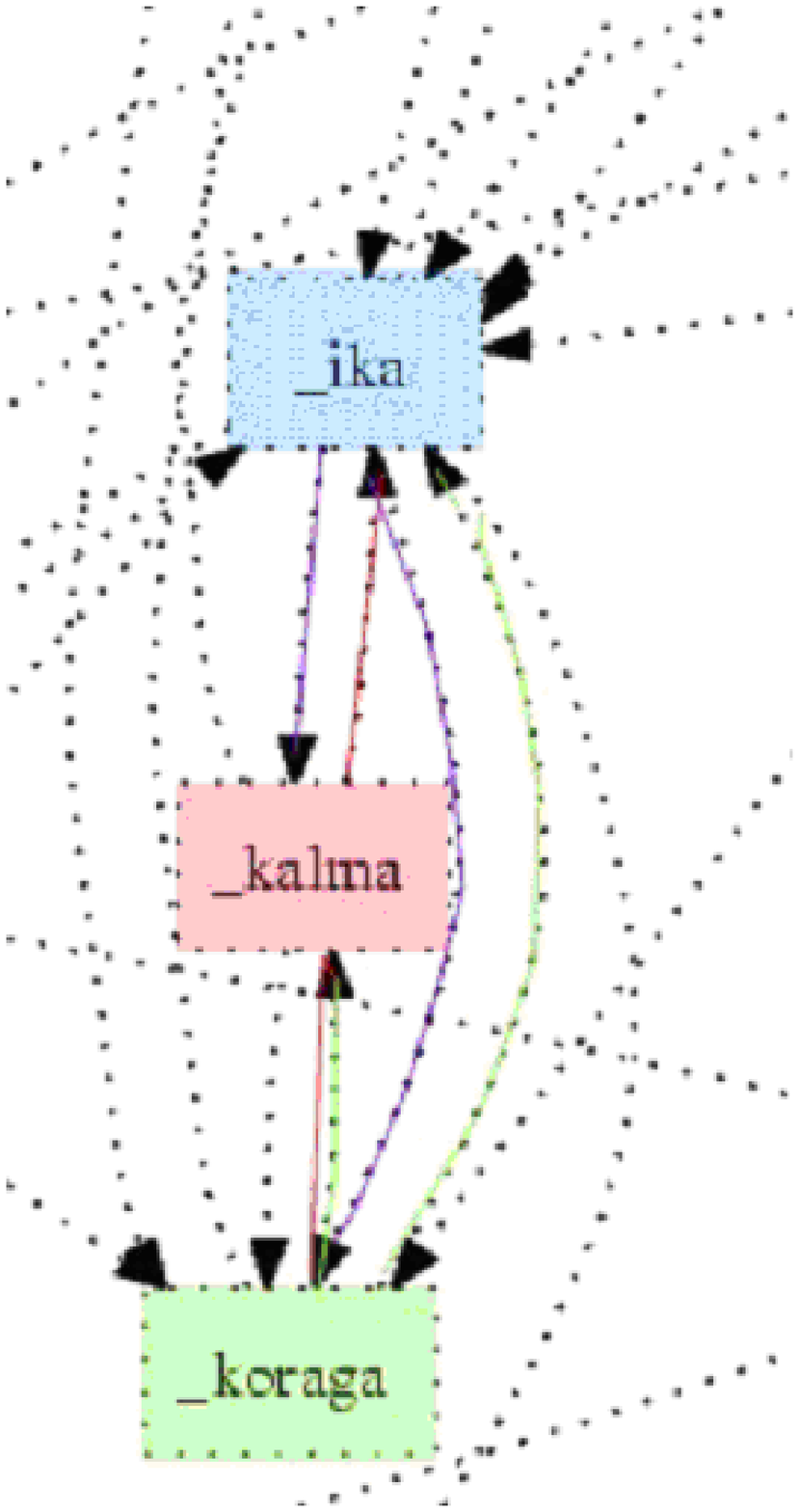}
\hspace{1cm}
\includegraphics[width=0.6\textwidth]{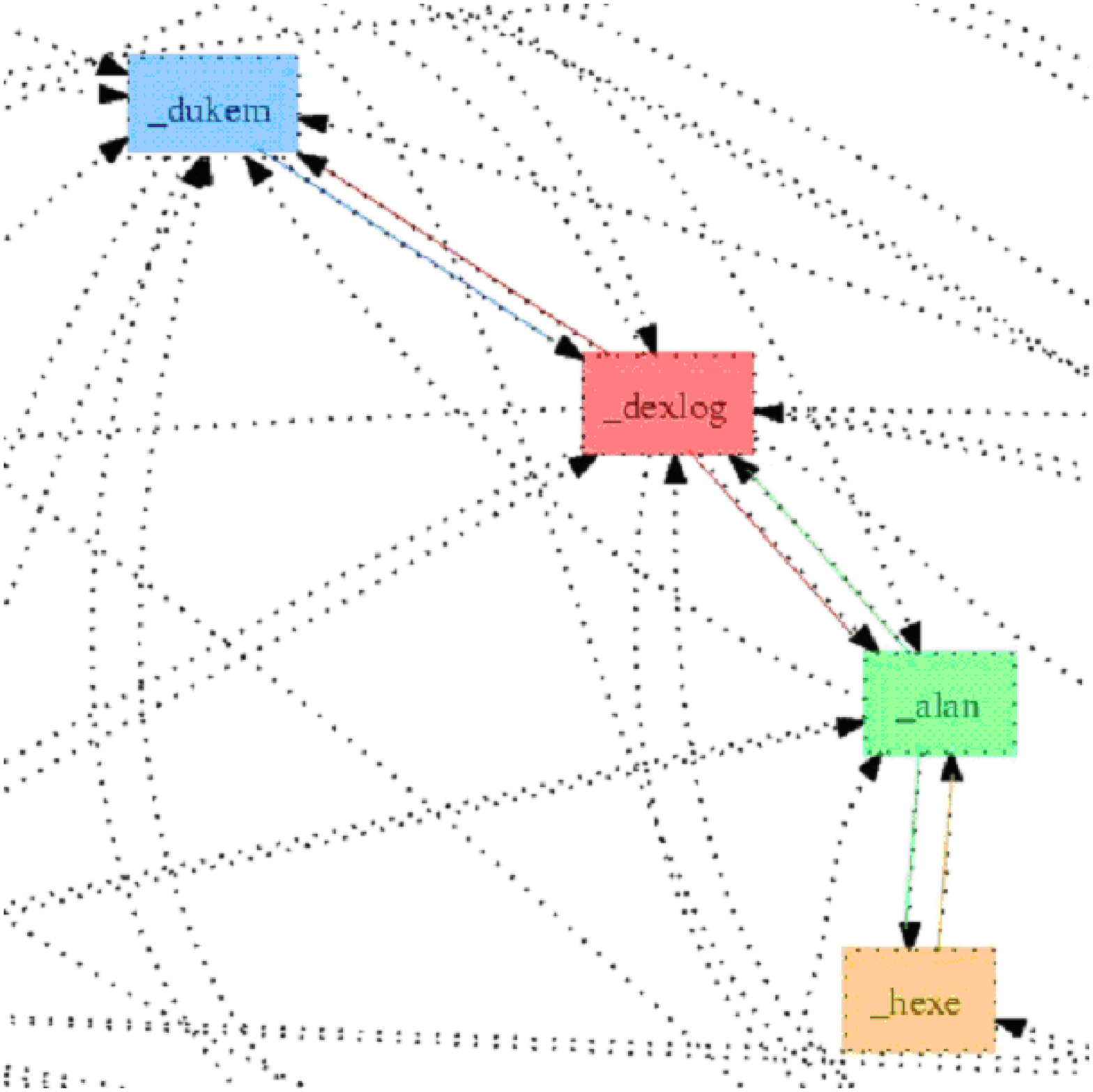}
\caption{Example of {\it triad} (left side) and chain of {\it diads} 
(right side) from jogger.pl}
\label{r6}
\end{figure}
       
In the left picture of Fig.~\ref{r6} 
the example of {\it triad} is shown --- three mutual choices from 
jogger.pl service. One can see that 
the positive  expansion of these blogs is small, despite relatively high sociometric status 
(with blog {\it kalma} having the smallest number of choices 
within the {\it triad}). Text analysis provides explanation of this --- all three 
blogs belong to one family,  a marriage with a 2 years old child. {\it koraga} is 
a blog describing events from child's life written from its 
``point of view'' by his mother, {\it kalma} is a weblog of its father while {\it ika} 
belongs to mother.
Right picture of Fig.~\ref{r6} shows the chain structure consisting of a number of {\it diads}.
Text analysis shows that these people are connected with historical internet 
portal {\it Histmag}. Such choice 
structure has been established despite large outlook differences.

\section{Summary}

Sparsity is the first apparent property of examined blogs networks. The highest observed
average vertex degree is $0.81$, that means that most of the vertices are not connected 
with others at all (about $90\%$).

$\gamma$ coefficients describing power--law of the decay of the vertex degree 
function is below $3.0$ in all examined services. That indicates that blogs are 
in fact scale--free networks \cite{barabasi, dorogovtsev}.

We don't observe notable increase of the average path length along with the 
increasing graph size. While the number of vertices of \textit{blog.onet.pl} and 
\textit{blog.gery.pl} services differs by order of magnitude, the difference between 
their average path lengths is only $0.84$. It can be observed that as the graph is 
growing, we don't need respectively longer paths to ``travel'' between its 
vertices. That  property is called \textit{small world} \cite{milgram}.

The proportion between strong and weak relations in cliques doesn't change with 
the size of the graph. However, small graphs are dominated by very dense (many 
connections) and loose cliques(no connections at all, isolated vertices). That 
contrast could be explained by saying that in smaller communities some people are 
very sociable, while others don't tend to ``connect'' with others at all. More 
balanced behaviour is rare.

In larger graphs, average cliquity is much greater (almost an order of magnitude) 
than in smaller ones, so we reckon that larger structures tend to help building stronger 
relations between their participants. In smaller structures the border between the
``liked'' and isolated ones is much stronger.

We also tried to implement a sociometric analysis method, a domain 
of microsociology, to analyse a large net of virtual interpersonal connections.
Although treating blog networks as such can be controversial, we believe 
that for the purpose of this analysis such interpretation can be proved valid.

In large groups it is possible to find some regular sociometric structures.
Structures described in this work were sociologically explainable despite vast 
differences of relationships among blog authors.

\section{Acknowledgments}

We'd like to thank our colleagues --- studens of Computer Science at Gda\'nsk University, 
who helped us with data processing: Krzysztof Treyderowski, 
Wojciech Glod, Marcin Jeremicz, Piotr Tadych, Lukasz Pasula, Lukasz Rolbiecki.

W.B., P.L. R.P. and D.M. are very grateful to organisers of the FENS meeting, 
professors R. Kutner and J. Holyst for their hospitality and subject inspiration.


\begin{thebibliography}{99}
\bibitem{sztompa} 1. Piotr Sztompka, \textit{Socjologia}, Wydawnictwo Znak, Krak\'{o}w 2002
\bibitem{a2} C.F. Nachmias, \textit{Research Methods in the Social Sciences}, Scientific American/St. Martin's College Publishing Group Inc. (1996)
\bibitem{a3} Oeconomicus: socjologia, \\ http://www.econom.pl/nauka/socjo5.php3 20.01.2005
\bibitem{a4} J. Szmatka, \textit{Male struktury spoleczne}, Warszawa 1989
\bibitem{a5} \textit{Male struktury spoleczne}, I. Machaj (ed.), Lublin 1998
\bibitem{a6} J. Turowski, \textit{Socjologia. Male struktury spoleczne}, Lublin 1993
\bibitem{a7} J. Brzezi\'{n}ski, \textit{Metodologia bada\'{n} psychologicznych}, Wydaw. Naukowe PWN, Warszawa 1999
\bibitem{barabasi} R. Albert and A.-L. Barabasi, \textit{Rev. Mod. Phys.} \textbf{74}, 47 (2002) 
\bibitem{dorogovtsev} S.N. Dorogovtsev and J.F.F. Mendes \textit{The shortest path to complex networks}, cond-mat/0404593
\bibitem{milgram} S. Milgram, \textit{Psych. Today} \textbf{2}, 60 (1967)
\end{thebibliography}
\end{document}